# Collective dynamics of microtubule-based 3D active fluids from single microtubules

Teagan E. Bate,[a] Edward J. Jarvis,[a] Megan E. Varney[a] and Kun-Ta Wu[*ab]

[a]Department of Physics, Worcester Polytechnic Institute, Worcester, Massachusetts 01609, USA
[b]The Martin Fisher School of Physics, Brandeis University, Waltham, Massachusetts 02454, USA

## Abstract

Self-organization of kinesin-driven, microtubule-based 3D active fluids relies on the collective dynamics of single microtubules. However, the connection between macroscopic fluid flows and microscopic motion of microtubules remains unclear. In this work, the motion of single microtubules was characterized by means of 2D gliding assays and compared with the flows of 3D active fluids. While the scales of the two systems differ by ~1,000×, both were driven by processive, non-processive or an equal mixture of both molecular motor proteins. To search for the dynamic correlation between both systems, the motor activities were tuned by varying temperature and ATP concentration, and the changes in both systems were compared. Motor processivity played an important role in active fluid flows but only when the fluids were nearly motionless; otherwise, flows were dominated by hydrodynamic resistance controlled by sample size. Furthermore, while the motors' thermal reaction led active fluids to flow faster with increasing temperature, such temperature dependence could be reversed by introducing temperature-varying depletants, emphasizing the potential role of the depletant in designing an active fluid's temperature response. The temperature response of active fluids was nearly immediate (≲10 sec). Such a characteristic enables active fluids to be controlled with a temperature switch. Overall, this work not only clarifies the role of temperature in active fluid activity, but also sheds light on the underlying principles of the relationship between the collective dynamics of active fluids and the dynamics of their constituent single microtubules.

## Introduction

Materials are comprised of building blocks held together by pairwise interactions that determine their material properties as well as their phases. In equilibrium, material states satisfy the rule of free energy minimization, whereas away from equilibrium, that rule is no longer applicable. A material stays out of equilibrium and is termed active matter when its constituents locally consume chemical energy to generate mechanical work. Active matter can consist of living organisms (such as bacteria and epithelial cells) or nonliving entities (such as colloids, grains and cytoskeletal filaments)[1-13]. Each constituent is motile and interacts with its neighbors[14]. These microscopic interactions accumulate to produce flocking, swarming or circulation at the macroscopic scale[15-22]. These unique characteristics of active matter have found application in optical devices, molecular shuttles and parallel computation, reflecting the feasibility and versatility of active matter applications[23-26]. However, development of these applications will be limited if knowledge of self-organization of active matter remains immature; therefore, there is a need to clarify the underlying principles of the dynamics of self-organization[27]. Therefore, this work correlated cumulative macroscopic dynamics with microscopic dynamics of constituents. The correlation revealed the role of microscopic dynamics in the macroscopic self-organization of active matter.

In this work, the active matter was kinesin-driven, microtubule-based 3D active fluids, due to their high tunability and their robust dynamics[28-31]. The system was comprised of three main components: microtubules, kinesin motor clusters and depletants (**Fig. 1A**). Depletants induced depletion forces that bundled microtubules[32]. The bundled antiparallel microtubules were bridged by motor clusters which consumed adenosine triphosphate (ATP) to slide apart microtubules, causing interfilament sliding. The sliding collectively led microtubule bundles to extend (extensile bundles, **Fig. 1B**) until buckling instability caused them to break[33]. The broken bundles recombined because of depletion forces, and the dynamics repeated. In the steady state, the microtubules developed into 3D active gels with dynamics consisting of repeating cycles of bundling, extending and buckling (**Fig. 1C**)[11]. The gels stirred surrounding fluids, inducing flows (active fluids) whose motion could be visualized by doping the fluids with tracer particles (**Movie S1**, ESI†). In this work, microtubule-based networks whose structures constantly rearranged are referred to as active gels, and the surrounding water driven by the motion of active gels as active fluids.

The flows of 3D active fluids originated from the motion of single microtubules whose dynamics could be characterized by monitoring a single 2D microtubule gliding on a kinesin-coated surface (gliding assay, **Fig. 1E&F**, **Movie S2**, ESI†)[34]. To characterize the collective dynamics from a single microtubule, mean speeds of 3D flows and 2D gliding were compared as motor dynamics were varied by varying temperature and ATP concentration[35-38]. To examine the role of the motor stepping mechanism in active fluid behaviors, the flows and gliding were driven with processive motors, non-processive motors and an equal mixture of both. Processive motors exerted a force on microtubules continuously whereas non-processive ones detached after each force application[31, 39-42]. Motor processivity dominated self-organization of active fluids only when the fluids were nearly still. At higher activity, active fluid dynamics were affected by hydrodynamic resistance[21, 43]. Also, while the motor's rising reaction rates with elevated temperatures led active fluids to flow faster when heated, such temperature responses of active fluids could be reversed by introduction of temperature-dependent depletants, suggesting the potential for depletant-based control of an active fluid's temperature dependence[44]. Finally, the temperature response of active fluids was not only reversible but also nearly immediate. Such an instant response enabled the design of active fluids whose activities were switchable, reinforcing the feasibility of controlling active matter *in situ* using temperature instead of electromagnetic control[4, 5, 45-49].

# Materials and Methods

## Polymerizing Microtubules

Active fluids were powered by kinesin-driven microtubules, comprised of α- and β-tubulin dimers. Tubulins were purified from bovine brains with three cycles of polymerization and depolymerization[50]. To polymerize microtubules, 8 mg/mL purified tubulins was mixed with 600 μM guanosine-5´[(α,β)-methyleno]triphosphate (GMPCPP, Jena Biosciences, NU-4056) and 1 mM dithiothreitol (DTT, Fisher Scientific, AC165680050) in microtubule buffer (M2B: 80 mM PIPES, 2 mM $MgCl_2$, 1 mM EGTA, pH 6.8)[51]. To image the polymerized microtubules, 3% of tubulin mixture was labeled with Alexa Fluor 647 (Invitrogen, A-20006). The mixture was incubated at 37°C for 30 min, then annealed at room temperature for 6 hr[11]. Polymerized microtubules were stored at –80°C until use.

## Synthesizing Kinesin Clusters

Tur active matter's power source originated from kinesin motors. The motors were expressed in the *Escherichia coli* derivative Rosetta 2 (DE3) pLysS cells (Novagen, 71403) transformed with DNA plasmids derived from the *Drosophila melanogaster* kinesin (DMK) gene. In this paper, plasmids were used that included the DMK's first 401 N-terminal DNA codons, which was long enough to permit the folding of the processive motor protein (K401)[40]. To determine the role of processivity in active gel dynamics, we employed another plasmid with the first 365 codons, which directs the expression of non-processive motors (K365, **Fig. 2A**)[41]. To drive active gels, the motors needed to be tagged with biotin and then clustered. Therefore, the motors were biotinylated at their N terminals (BIO) and then tagged with a histidine tag (H6), enabling purification of the motors using immobilized metal ion affinity chromatography (IMAC) with gravity nickel columns (GE Healthcare, 11003399)[52]. Thus, processive motors consisted of K401-BIO-H6, and non-processive ones of K365-BIO-H6[40, 41]. To cluster either biotinylated motor, streptavidin tetramers were used as the clustering agent; 1.8 μM streptavidin (Invitrogen, S-888) was mixed with M2B containing 120 μM DTT and either 1.5 μM processive motors or 5.4 μM non-processive motors, followed by 30 min incubation at 4°C[30, 31]. The clustered motors were aliquoted and stored at –80°C.

## Preparing Kinesin-Driven, Microtubule-Based Active Fluids

Active fluids were driven by ~0.1% active gels, which were comprised mainly of microtubules, kinesin motors and depletants (**Fig. 1A**). To synthesize active gels, 1.3 mg/mL microtubules was mixed with 120 nM motor clusters in high salt M2B (M2B + 3.9 mM $MgCl_2$). To bundle microtubules, either 0.8% polyethylene glycol (PEG 20k, Sigma, 81300) or 2% pluronic F127 (Sigma, P2443) was added as a depletant. The gel's activity was supported by motors stepping along microtubules from minus to plus ends[53]. Each step consumed adenosine triphosphate (ATP), which was converted to adenosine diphosphate (ADP); therefore, the stepping was fueled with 1.4 mM ATP (Sigma, A8937)[38]. However, the consumption of ATP decreased its concentration and slowed the stepping rate[37, 38]. To prevent such deceleration, ATP concentration was maintained by reverting each stepping-induced ADP back to ATP with 2.8% v/v stock pyruvate kinase/lactate dehydrogenase (PK/LDH, Sigma, P-0294). To support the reversion, the enzymes were fed with 26 mM phosphenol pyruvate (PEP, BeanTown Chemical, 129745)[11, 54]. In this work, samples were imaged with fluorescent microscopy whose quality decreased with time due to photobleaching effects. To reduce these effects, the active gel mixtures were doped with 2 mM Trolox (Sigma, 238813) and oxygen-scavenging enzymes, including 0.038 mg/mL catalase (Sigma, C40) and 0.22 mg/mL glucose oxidase (Sigma, G2133), fed with 3.3 mg/mL glucose (Sigma, G7528). The system was based on proteins, which needed to be stabilized with reducing agents; therefore, the mixtures were doped with 5.5 mM DTT. Finally, active fluids were transparent; to track their motion, ~0.0004% v/v Alexa 488-labeled 3 μm polystyrene microspheres (Polyscience, 18861) were added as tracers (**Fig. 1C** and **Movie S1**, ESI†). Mixing the aforementioned ingredients completed the synthesis of the active fluids. However, different samples of active fluids prepared in the same way could behave differently due to pipetting-induced uncertainties. To reduce such uncertainties and enhance experimental reproducibility, a batch of the above mixture without microtubules (premixture) was prepared[30]. The premixture was aliquoted and stored at –80°C. Before each experiment, active fluids were prepared by mixing one aliquot of premixture with microtubules. After the mixing, the sample was activated: motors started to consume ATP while stepping along microtubules.

Characterizing Active Fluid Activity

The activated sample created turbulent-like mixing flows[11, 21]. To observe the flows microscopically, the samples were loaded into a polyacrylamide-coated glass flow cell (~20×5×0.1 mm³ = 10 μL), and sealed with UV glue (Norland Products, NOA81)[31, 55]. Once sealed in flow cells, active fluids were imaged with epifluorescent microscopy: Alexa 647-labeled microtubules were imaged with a Cy5 filter cube (Excitation: 618–650 nm, Emission: 670–698 nm, Semrock, 96376), and Alexa 488-labeled tracers were imaged with a green fluorescent protein (GFP) filter cube (Excitation: 440–466 nm, Emission: 525–550 nm, Semrock, 96372) (**Fig. 1C**). To characterize active fluids' activity, tracers were imaged sequentially, and their trajectories were tracked with the Lagrangian tracking algorithm[56]. The trajectories $r_i(t)$ revealed the history of their instant speeds $v_i(t) \equiv dr_i(t)/dt$, which determined the evolution of the fluid's mean flow speed $\bar{v}(t) \equiv \langle v_i(t) \rangle_i$. At 20°C, active fluids flowed at ~6 μm/s for hours, until ATP and PEP were depleted (**Fig. 1D**)[21]. At 30–36°C, the flows sped up to ~10 μm/s but slowed down gradually because the bundles coarsened over time which built up friction to hamper bundle extension (Coarsening of Microtubule Bundles in ESI†). At 40°C, the flows quickly decayed due to malfunctioning kinesin motor clusters (Temperature-Induced Malfunction for Kinesin Clusters in ESI†). At 10°C, the flows decelerated nearly to a stop (~0 μm/s) due to depolymerization of GMPCPP-stabilized microtubules (Characterization of Microtubule Depolymerization in ESI†). The variation of fluid mean speeds depended on bundle coarsening and stability of microtubules and kinesin motor clusters, all of which varied with temperature.

Separating Active Kinesin Motor Proteins

To ensure the expressed motor proteins were capable of hydrolyzing ATP to produce mechanical work when in contact with microtubules, 0.7 mg/mL kinesin was incubated with 2 mg/mL microtubules, 3.57 mM DTT and 1.4 mM ATP at ~23°C for 30 min. Functioning motors hydrolyzed ATP, then detached from microtubules after ≳100 steps[39, 57], while malfunctioning motors bound to microtubules without detaching. After incubation, the functioning and malfunctioning motors were separated by 15-min, 100,000 $g$ centrifugation at 37°C. The malfunctioning motors sedimented with microtubules while active motors remained suspended. The suspended active motors were then aliquoted and stored at –80°C.

Gliding Microtubules on Kinesin-Coated Surfaces

To evaluate the dynamics of 3D microtubule-based active fluids from dynamics at the scale of single microtubules, individual microtubules were placed on kinesin-coated surfaces (gliding assay, **Fig. 1E**). To prepare such surfaces, cleaned glass coverslips and slides were etched by 10 min of sonication in 100 mM potassium hydroxide (KOH). The etched glasses were incubated in 1 mg/mL PEG silane (Laysan Bio, MPEG-SIL-2000-1g) and 39 μg/mL biotinylated PEG silane (Laysan Bio, Biotin-PEG-SIL-3400-1g) in ethanol at 70°C for 5 min. The former coated the glass surface with PEG, preventing proteins from sticking to the bare glass surfaces; the latter activated the surfaces with biotin, which was further conjugated to streptavidin through 30 min incubation in 1 mg/mL streptavidin in M2B at 4°C. After incubation, the unbound streptavidin was washed out with M2B. The coated streptavidin was subsequently conjugated to biotinylated kinesin motors by 20 min incubation in ~0.5 mg/mL clarified biotinylated kinesin motor proteins in M2B at 4°C. After incubation, the free kinesin motors were washed out with M2B, leaving the glass surfaces coated with kinesin motors.

The coated motors were then bound to microtubules through 2 min incubation of the coated slides in 50 μg/mL Alexa 647-labeled microtubules in M2B at ~23°C. After incubation, the unbound microtubules were washed out with M2B. To activate microtubule gliding, slides were loaded with ATP solution (5.6 mM DTT, 3.33 mg/mL glucose, 255 μg/mL glucose oxidase, 39 μg/mL catalase, 2 mM Trolox, 1.42 mM ATP in high salt M2B), enabling the kinesin motors to hydrolyze ATP while driving microtubule gliding (**Fig. 1E**). To promote surface adsorption of microtubules, 0.4% PEG was added to the solution. PEG also induced depletion among microtubules; however, no bundle formation was observed, suggesting that 0.4% PEG provided insufficient depletion to bundle microtubules in this gliding assay[9, 10]. To monitor gliding, microtubules were sequentially imaged with fluorescence microscopy using Cy5 filter cubes and a 60× objective (Nikon, CFI Plan Fluor 60X NA 0.85, MRH00602) for 5 min (**Fig. 1F** and **Movie S2**, ESI†). The sequential images allowed tracking of the gliding of each microtubule with the Lagrangian tracking algorithm[56]. The trajectories revealed gliding speeds *vs.* time, which were averaged over different microtubules to determine mean gliding speeds. At 20°C, microtubules glided on K401-coated surfaces at ~0.3 μm/s (**Fig. 1G**).

## Controlling Sample Temperature

To tune motor activity locally, sample temperature was varied. The sample was cooled or heated through contact with a peltier (TE Technology, CH-109-1.4-1.5, **Fig. S4A**). Peltier performance depended on the direction and amplitude of the applied direct current, which was regulated by a temperature controller (TE Technology, TC-720)[58]. The controller read the sample temperature through a thermal sensor (TE Technology, MP-3022), and the sample temperature was compared with the target temperature to determine the applied current through a proportional-integral-derivative (PID) algorithm[59]. The controller was connected to a computer for recording sample temperatures. The recorded temperatures showed a fluctuation of ≲0.3°C, demonstrating the stability of the temperature control (**Fig. S4B**). Such controllability enabled characterizing the sample's response to temperature.

## Tuning Kinesin-Driven Dynamics with Temperature and ATP Concentration

In this work, kinesin dynamics were tuned with temperature and ATP concentration[35, 38]. First, mean speeds of K401-driven active fluids and microtubule gliding were measured as a function of time between 10 and 40°C. In the active fluid system, the change in mean speed depended on temperature (**Fig. 1D**). At 20°C, the mean speed was nearly constant, whereas above or below 20°C, the mean speed decayed with time. To characterize system activity, mean speeds were averaged within a common window, between $t = 1$ and 2 hr. During the first hour of data acquisition, mean speeds could change drastically due to the malfunction of kinesin motor clusters at higher temperatures and microtubule depolymerization at lower temperatures (Characterization of Microtubule Depolymerization and Temperature-Induced Malfunction for Kinesin Clusters in ESI†). Because of this, the beginning of the averaging window was set to 1 hr. After the first hour the mean speed of the system still exhibited gradual change over time due to the coarsening of microtubule bundles (Coarsening of Microtubule Bundles in ESI†). Coarsening was a common issue in this active fluid system[31]; therefore, averaging for a 1 hr period ensured analysis with sufficient statistics while reducing the influence of the coarsening effect. On the other hand, the mean speeds in the microtubule gliding assay were nearly time independent (**Fig. 1G**). The choice of averaging windows did not affect analysis results in the gliding assay. However, the image acquisition setup had an issue of slow focal plane drift. The drift was only a few microns, but it blurred images of microtubules gliding on the glass surface after 10–30 min of image acquisition, depending on ambient temperature. Analysis on blurry microtubule images was not reliable. Therefore, mean speeds of microtubule gliding were averaged during the first 5 min ($t = 0$–5 min), to ensure analysis with sufficient statistics while reducing the out-of-focus issue. Averaging mean speeds in active fluids and gliding systems revealed their temperature dependence (**Fig. 2B**).

To illustrate the role of motor processivity, measurements were repeated with K401 replaced by K365 (**Fig. 2C**). The former stepped along microtubules continuously; the latter detached after each step (**Fig. 2A**)[40, 41]. To characterize how the different stepping mechanisms interacted to impact system dynamics, the measurements were repeated with both types of motors mixed in equal amounts. In active fluids, motor clusters were replaced with equal concentrations of K401 and K365 clusters (60 nM); in the gliding assay, the coated motors were 50% K401 and 50% K365 (**Fig. 2D**).

Lowering ATP concentrations slows down motor stepping, leading to slower dynamics in active fluid and gliding systems[11, 38]. In this work, ATP concentration was also varied over the ranges between 45 and 1,400 mM while keeping the temperature at 20°C (**Fig. 3**). These experiments not only characterized how systems driven by different types of motors responded to temperature and ATP concentration, but also enabled the search for a dynamic connection between the two distinct but related systems by correlating mean speeds of 3D active fluid flows and 2D microtubule gliding (**Fig. 4**).

## Measuring Depletant Sizes with Dynamic Light Scattering

The dynamics of the active fluids relied not only on the interfilament sliding driven by kinesin motor clusters but also on the microtubule bundling induced by depletants (**Fig. 1A**). Characterizing the role of depletants in these active fluids required determination of depletant sizes. However, the depletants were <100 nm, which could not be imaged with optical microscopy. Instead, their sizes were measured with dynamic light scattering (Malvern Panalytical, Zetasizer NANO S90). To perform such a measurement, the depletants were suspended in high salt M2B loaded into a glass cuvette (Starna, 1-G-10). The cuvette center was exposed to a laser of wavelength $\lambda$ = 633 nm. The laser beam was scattered by the suspended depletant. The scattered light was collected by a detector angled from the beam by $\theta$ = 7° (**Fig. 5A**). The collected light intensities $I(t)$ fluctuated because of depletant diffusion. Therefore their autocorrelation functions $g(\tau) \equiv \langle I(t) \rangle_t \langle I(t+\tau) \rangle_t / \langle I(t) \rangle_t^2$ decayed exponentially with a decay rate $\Gamma$ proportional to the depletant's diffusion coefficient $D$:

$$\Gamma = \frac{32\pi^2 n^2 \sin^2(\frac{\theta}{2})}{\lambda^2} D,$$

where $n$ was the refractive index of the solution (**Fig. 5B** inset)[60, 61]. Here, the solution refractive index was assumed to be similar to that of water, $n \approx n_w = 1.33$. The diffusion was tied to suspension sizes: $D = k_B T / 3\pi\eta d$, where $k_B$ was the Boltzmann constant, $\eta$ was water viscosity and $d$ was suspension diameter (Einstein relation)[62]. Therefore, through measuring the decay rate, the depletant sizes were determined: PEG, ~11 nm; pluronic F127, ~9 nm (**Fig. 5B**). To examine the impact of temperature on depletants, these sizes were measured from 10 to 40°C, revealing that the size of PEG was nearly invariant. In contrast, the size of pluronic increased rapidly from 20 to 25°C and saturated at ~22 nm at 25°C (**Fig. 5B**). Such a change implied micelle formation, and the temperature dependence of pluronic-induced depletion[44, 63]. To examine the impact of temperature-dependent depletion in active fluids, PEG was replaced with pluronic and the time-averaged mean speeds were measured from 10 to 40°C (**Fig. 5C**).

# Results and Discussion

Self-organization of active fluids relied on cumulative dynamics of pairs of microtubules sliding apart (**Fig. 1A–C**)[11]. However, since these microtubules were ~1 μm long, monitoring individual interfilament sliding in 3D bulk was challenging. Instead, the sliding dynamics were assumed to be similar to those of single microtubule gliding on a 2D motor-coated surface (gliding assay, **Fig. 1E&F**)[10, 34]. In this work, gliding assay was used as a model system to characterize the motility of single microtubules driven by kinesin motors. Because activities of both systems originated from kinesin motor proteins that consume ATP to generate mechanical work whose energy transducing rates could be tuned with ATP concentration and temperature, microtubule gliding speeds were compared with flow speeds of active fluids to reveal the dynamic relationship between the two systems.

## Temperature Dependence of Active Fluids and Microtubule Gliding Assay Follows the Arrhenius Law

To characterize the dynamic connection between the two systems, their temperatures were varied between 10 and 40°C while their mean speeds were measured (**Fig. 2B**). Kinesin-driven systems have been known to be temperature dependent as kinesin is a mechanochemical enzyme whose reaction is diffusion-limited; additionally, kinesin stepping behaviors require diffusion and are therefore temperature-dependent[39, 64, 65]. This temperature dependence has been described using the Arrhenius Law: reaction rate $k \propto e^{-E_a/RT}$, where $R$ = 8.31 J/mol K is the gas constant, $T$ is temperature, and $E_a$ is activation energy[35, 66, 67]. The activation energy represents the reaction barrier; a larger activation energy implies a more temperature-sensitive reaction. The activation energy has been measured in the kinesin-driven gliding assay system to be 10–100 kJ/mol[35, 36, 45]. The variation in previously measured activation energies depended on the specific motor mechanism. To illustrate how the motor mechanism impacts the activation energy in both of our systems, active fluids and microtubule gliding were driven with three different motor systems: processive motors (K401), non-processive motors (K365), and an equal mixture of both (50% K401 and 50% K365, termed mixed-motor). K401 is double-headed processive motors that step along microtubule protofilaments hand-over-hand without detaching, whereas K365 is single-headed non-processive motors which detach after each step (**Fig. 2A**)[31, 41, 42]. A single processive kinesin can propel a microtubule in a gliding assay, whereas 5-6 motors are required if the kinesin is non-processive[42].

To measure activation energies of the three different motor systems, the mean speeds of microtubule gliding were measured between 10 and 40°C, followed by fitting the mean speed *vs.* temperature to the Arrhenius equation: $v =$

$Ae^{-E_a/RT}$ with $A$ and $E_a$ as fitting parameters. Only the data between 16 and 36°C was used in these fittings, for the following reasons. First, the GMPCPP-stabilized microtubules depolymerized below 16°C (Characterization of Microtubule Depolymerization in ESI†). It seemed likely that data confounded by depolymerization (< 16°C) would not exhibit the same dynamics as data derived from stable microtubules (≥ 16°C). Second, motors used in these experiments that were pre-cooked at $T > 36$°C did not develop the same dynamics with respect to microtubules, indicating the malfunctioning of the motor clusters (Temperature-Induced Malfunction for Kinesin Clusters in ESI†)[35]. For the above reasons, these microtubule-kinesin systems were deemed to be unstable at $T < 16$°C and $T > 36$°C; therefore, the data in these ranges was not be mixed nor analyzed with data derived from systems in which microtubules and kinesin clusters were both stable (16–36°C). As such, in this temperature range, the activation energies were measured in processive, non-processive and mixed-motor systems to be 43, 55 and 28 kJ/mol, respectively (magenta, cyan and light green in **Fig. 2B–D**). Two observations resulted from these measurements. First, the activation energy of non-processive motor system was higher than that of the processive system, possibly because non-processive motors detached from microtubules after each step. After detaching, the motors were not able to propel the microtubules until they had diffused back to the microtubules. The diffusion process was temperature-dependent; therefore, K365-driven dynamics were more temperature-sensitive, which led to a higher activation energy. Second, the mixed-motor system had the lowest activation energy. The result is counterintuitive as one might expect that mixing motors should not alter the activation energy. However, the measurement results suggested that cooperation between processive and non-processive motors led to less temperature sensitive dynamics (lower $E_a$). Gaining insights into the impact of motor cooperation effects requires further studies which may involve computation and models considering each motor's association and dissociation rates along with their stepping rates and duty cycles[34].

Followed characterization of the gliding assay, we shifted the focus to the active fluid system. In this system, mean speeds of active fluid flows were measured between 10 and 40°C, and activation energy was determined by fitting the mean speed *vs.* temperature (16–36°C) to the Arrhenius equation. The activation energies in processive, non-processive and mixed-motor systems were 22, 27, and 13 kJ/mol, respectively (red, blue and dark green in **Fig. 2B–D**). Three observations resulted from these measurements. First, mixed-motor system had the lowest activation energy of the three, consistent with the results of the gliding assay and demonstrating the need for further studies about cooperation between the two types of motors. Second, the activation energies of processive and non-processive motor systems were nearly indistinguishable (within fitting errors, **Fig. 2B–D**). Such similarity indicates that the motor processivity does not play an important role in the active fluid system. This result is contrary to the results in the gliding assay, but consistent with previous studies by Chandrakar *et al.*, which showed that active fluids with different nanoscopic mechanisms develop similar macroscopic dynamics[31]. Finally, activation energies in active fluid systems were lower than in the corresponding gliding assay systems that used the same motors. The underlying mechanism that causes such a reduction is unclear, but it is possible that the mechanism involves the intervention of temperature-dependent hydrodynamic viscosity. Viscosity plays an important role in self-organization of fluids, and within the temperature range that we explored here (16–36°C), the viscosity changes by ~60%[68]. Therefore, to better interpret the dynamics of active fluids requires development of a more sophisticated computational model that considers not only microtubule-associated motor dynamics but also hydrodynamics involving fluid viscosity.

## ATP Dependence of Active Fluids and Microtubule Gliding Assay are Captured by Michaelis–Menten Kinetics

Motor activities are affected not only by temperature but also by ATP concentration[35, 38]. Varying [ATP] changes motor hydrolysis rate linearly until it reaches the saturation maximum (Michaelis–Menten kinetics)[69]. Such a kinetic theory has been applied to gliding assays[37, 38, 70]:

$$\text{Mean speed, } v = v_m[\text{ATP}]/(c_m + [\text{ATP}]),$$

where $v_m$ is the saturation maximum speed and $c_m$ is the ATP concentration leading to half saturation, $v_m/2$. In numerous previous studies, the measured $v_m$ ranged from 100 to 1,000 nm/s[35, 41, 71, 72]. To measure $v_m$ in the gliding assay, mean speed of microtubule gliding was measured as a function of [ATP] between 45 and 1,400 μM, followed by fitting the mean speed *vs.* [ATP] to Michaelis–Menten kinetics with $c_m$ and $v_m$ as fitting parameters (**Fig. 3A–C**). In processive, non-processive and mixed-motor systems, the measured $v_m$ was 360, 120 and 270 nm/s, respectively (**Fig. 3D**). Two observations resulted from these measurements. First, the $v_m$ for non-processive motors was lower than for processive motors, which was consistent with previous studies showing that the lower speed of the non-processive motors was due to their lower duty cycles[41]. Second, the $v_m$ in mixed-motor system was nearly the average of that of the other two systems, suggesting that each type of motor proportionately contributed driving force to propel microtubules. This observation demonstrated the feasibility of tuning microtubule gliding speeds based on the proportion of surface coating of fast and slow motors.

To examine how $v_m$ was impacted when motor dynamics were accumulated in millimeter-scale active fluid flows, mean speeds of active fluid flows for [ATP] = 45–1,400 µM were measured, followed by fitting the mean speed *vs.* [ATP] to Michaelis–Menten kinetics (**Fig. 3A–C**). In processive, non-processive and mixed-motor systems, $v_m$ was measured as 6.2, 3.4 and 4.8 µm/s, respectively (**Fig. 3D**). Three observations resulted from these measurements. First, the non-processive motor system had a lower $v_m$ than the processive one. Second, $v_m$ of the mixed-motor system was the average of the other two. Both results were consistent with those from the gliding assay, demonstrating that dynamics in active fluids and the gliding assay were connected. Third, $v_m$ in active fluids was ~10× faster than in the gliding assay. This difference suggests that the dynamics in active fluids result from the collected dynamics of millions of motors at nanoscale.

## Cross-Comparing 3D Active Fluid Flows with 2D Microtubule Gliding

To gain deeper insight into the collective dynamics of active fluids, the mean speed of active fluids ($v_{3D}$) and microtubule gliding ($v_{2D}$) were cross-compared at the same temperature and ATP concentration (**Fig. 4**). Gliding exhibited the dynamics of single microtubules at the micron scale, whereas fluids exhibited the collective dynamics of multiple microtubules at the millimeter scale. The comparison revealed the dynamic correlation between the two systems. In the fast gliding regime ($v_{2D} \gtrsim 0.2$ µm/s), the $v_{3D}$-$v_{2D}$ correlation was nearly indistinguishable among the explored systems (processive, non-processive and mixed-motor), consistent with the previous observation that motor processivity did not significantly impact active fluid dynamics. However, this similarity faded at gliding speeds $v_{2D} \lesssim 0.2$ µm/s. Decreasing $v_{2D}$ by ~70% decelerated the corresponding $v_{3D}$ by ~250% in processive motor-associated systems (red and green), whereas in non-processive motor systems (blue), a similar variation in $v_{2D}$ (~70%) led to a smaller change in $v_{3D}$ (~80%). Such an observation came mainly from the experiments with decreasing [ATP] (**Fig. 3**). Low [ATP] has been demonstrated to cause processive motors to pause between steps[73-78]. The pausing motors have been shown to firmly attach to microtubules[79], implying that the pausing motor clusters can crosslink microtubule networks, thus mechanically increasing the hindering load for each motor[34, 80]. The load decreases motor stepping rate and consequently inhibits microtubule network activities, which leads to slower flows of active fluids[75, 81]. In contrast, in the non-processive system, motors detach after each step and therefore have a lower probability of crosslinking microtubules[31, 41], resulting in faster active fluid flows in the range $v_{2D} \lesssim 0.2$ µm/s. This cross-comparison broadened the previous finding regarding motor processivity in active fluids: motor processivity did not play an important role in dynamics of active fluids only when the corresponding gliding assay was in the fast regime ($v_{2D} \gtrsim 0.2$ µm/s). In the slow ($v_{2D} \lesssim 0.2$ µm/s) or low [ATP] regime, the dynamics were influenced by motor processivity.

Fluid dynamics are governed by the Navier-Stokes equation, and the role of hydrodynamics in active fluids remains to be elucidated[82]. To advance understanding of such a role, measurements of $v_{3D}$-$v_{2D}$ were repeated with K401-driven active fluids but in a flow cell 7× thicker (~20×5×0.7 mm³ = 70 µL). The thicker cell induced lower hydrodynamic resistance and therefore supported faster flows (**Fig. 4** inset)[21, 43]. Moreover, the larger cell did not impact the $v_{3D}$-$v_{2D}$ correlation except that it shifted $v_{3D}$ upward by ~3-fold (arrow in the inset). The shift indicated that in active fluids, hydrodynamics played the role of scaling, consistent with previous studies showing that flow profiles of confinement-induced coherent flows scaled with confinement size[21].

## Reversing Temperature Dependence of Kinesin-Driven Active Fluids

Kinesin motor-driven active fluids were influenced by temperature according to the Arrhenius Law (**Fig. 2B–D**)[35, 66, 67]. However, to what extent such dependence in active fluids could be controlled remained unexplored. To seek an alternative for tuning the temperature dependence of active fluids, the temperature-independent depletant PEG was replaced with temperature-dependent pluronic F127 (**Fig. 5B**). Measurements of mean speed *vs.* temperature showed that, as with PEG, F127-doped active fluids moved faster as temperature increased from 10°C; however, above 20°C, fluids decelerated rapidly (**Fig. 5C**). Such deceleration was caused by two factors. First, increasing temperatures above 20°C triggered the formation of micelles (**Fig. 5B**)[44]. Micellization consumed pluronic monomers, reducing depletion between pairs of microtubules. Lack of sufficient depletion suppressed microtubule bundling as well as interfilament sliding; thus fluid activity was inhibited. Second, formation of pluronic micelles created hydrophobic cores which potentially bound to kinesin, causing denaturing[63, 83]. The denatured motors could not support fluid activity. Verifying these arguments requires additional studies, such as characterizing depletion change during pluronic micellization and clarifying the impact of hydrophobic cores of pluronic micelles on kinesin motor proteins. Nevertheless, this work demonstrates a potential means to reverse the temperature dependence of kinesin-driven active fluids by harnessing the temperature-dependent characteristics of pluronics.

### Tuning Active Fluid Activities *in Situ*

Supporting active fluid flows require both motor and depletion forces (**Fig. 1A**). The depletion force can be regulated by changing concentrations of depletants suspended in fluids; therefore, changing depletants requires either preparing another sample or flowing fresh active fluids into microfluidic devices[84]. Both methods erase the original materials and replace with new fluids which wipe out the original sample's history. Preserving the materials and sample history while tuning fluid activity locally can be achieved by varying motor dynamics with temperature. While motor-driven gliding has been demonstrated to respond to temperature reversibly and instantly, to what degree such reversibility along with the fast response can be derived to 1,000× larger active fluids remains unclear[45]. Here, the temperature of active fluids was alternated between 20 and 30°C every 30 min. The alternation accelerated and decelerated motor dynamics periodically, which led active fluids to flow faster and slower accordingly (**Fig. 6**, **Movie S3**, ESI†). Moreover, active fluids responded to temperature changes nearly instantly ($\lesssim 10$ sec). Such an instant response demonstrated the feasibility of designing active fluids with switchable flows, paving the way for the creation of valveless microfluidic devices.

## Conclusions

We have cross-compared 3D flows of active fluids with 2D microtubule gliding when both systems are driven by processive, non-processive and an equal mixture of both molecular motors. The comparison highlighted the roles of hydrodynamics and motor processivity. The hydrodynamic influence was long ranged (~1 millimeter), scaling the resulting dynamics, whereas motor processivity-induced influence was short ranged (~1 micron), controlling the dynamics in slow flow regimes. The results implied that the ranges of influences dominated their roles in the collective dynamics of molecular motor-driven, cytoskeleton-based active matter. However, to what extent this conclusion can be generalized to generic active matter remains an open question. Such a question can be explored by adapting this method of comparison, which connects microscopic with macroscopic dynamics, for advancing the knowledge of collective dynamics of active matter.

## Conflicts of interests

There are no conflicts to declare.

## Author Contributions

T.E.B. and E.J.J. performed the research; T.E.B., E.J.J. and K.T.W. designed the experiments; T.E.B. characterized temperature dependence of 3D active fluids; E.J.J. acquired data in the gliding assay; M.E.V. performed dynamic light scattering; T.E.B., E.J.J., M.E.V. and K.T.W. analyzed the data; T.E.B., E.J.J., and K.T.W. wrote the manuscript. All authors have reviewed the manuscript.

## Acknowledgements

Plasmids of K401-BCCP-H6 and K365-BCCP-H6 were gifts from Zvonimir Dogic. We thank Arne Gericke for the use of his Zetasizer NANO S90 in our light-scattering measurements. This research was performed using computational resources supported by the Academic & Research Computing group at Worcester Polytechnic Institute. We acknowledge Brandeis MRSEC (NSF-MRSEC-1420382) for use of the Biological Materials Facility (BMF).

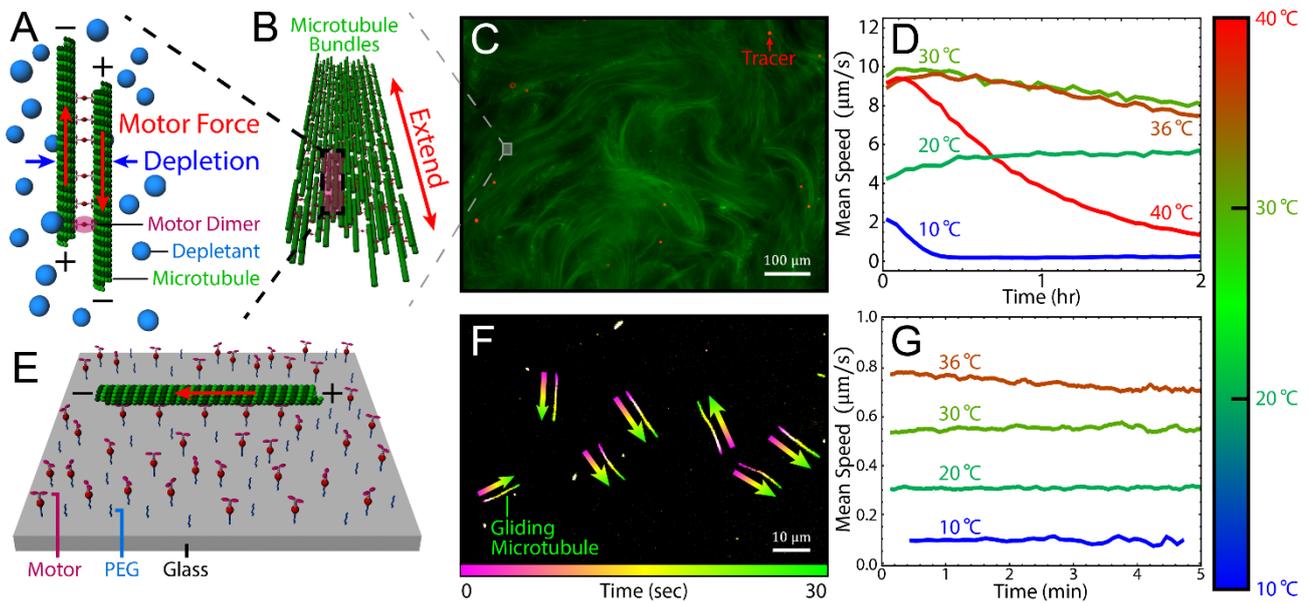

**Fig. 1**: Comparison of Temperature-Dependent Dynamics of 3D Microtubule-Based Active Fluids with 2D Microtubule Gliding. (A) Pairs of microtubules were brought together by depletant-induced attraction (depletion). The bound microtubules were then bridged by motor clusters that slide apart antiparallel microtubules, causing interfilament sliding. (B) The sliding resulted in the extension of microtubule bundles (extensile bundles). (C) The collective dynamics of extensile bundles constituted active microtubule networks (green filaments), which constantly rearranged their structure while stirring surrounding aqueous fluids whose motion was revealed with tracers (red dots). (D) Tracer trajectories were used to determine fluid mean speeds, which depended on temperature. (E) Schematic of a microtubule gliding on a motor-coated surface (gliding assay). (F) Stacking sequential images of gliding microtubules colored according to a time color bar revealed gliding trajectories, directions (magenta to green) and speeds (trajectory lengths). (G) The mean speeds of microtubule gliding were nearly invariant for 5 min but the speeds increased with temperature.

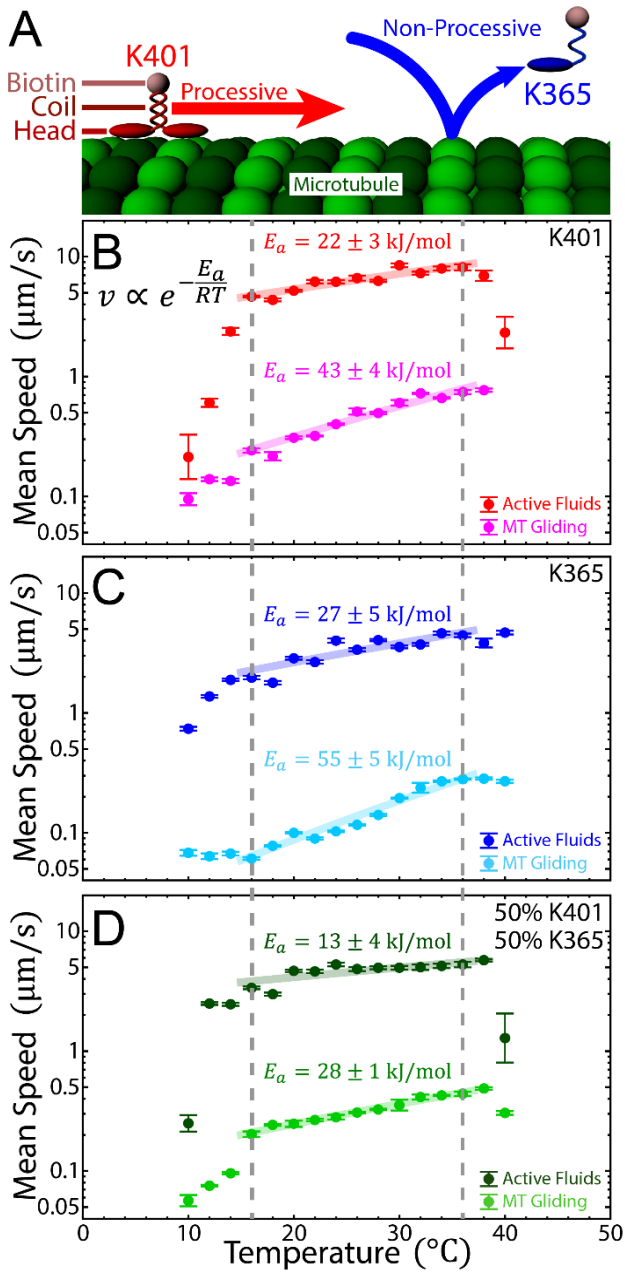

**Fig. 2**: Temperature Response of Active Fluids and Microtubule Gliding Driven by Processive and Non-Processive Motors. (A) K401 motors walked along microtubules continuously without detaching (processive) whereas K365 motors detached after each step (non-processive). Both types of motors were biotinylated (pink spheres) to enable conjugation to streptavidin. Streptavidin conjugation enabled the formation of motor clusters in active fluids, as well as motor-coating of surfaces in a gliding assay (**Fig. 1A&E**). (B–D) Mean speeds of active fluid flows and microtubule gliding as a function of temperature. Decreasing the temperature below 14°C destabilized microtubules (Characterization of Microtubule Depolymerization in ESI†); flows and gliding were suppressed. Heating above 38°C inhibited K401-associated activities, implying occurrences of malfunction in K401 motor clusters, whereas K365-driven systems remained active, demonstrating a stronger resistance to heat (Temperature-Induced Malfunction for Kinesin Clusters in ESI†). Increasing the temperature from 16 to 36°C (between dashed gray lines) accelerated flows and gliding driven by K401 and K365 but not for an equal mixture of both (50% K401 and 50% K365). Active fluids driven by both types of motors had temperature-insensitive flows. Error bars represent the standard deviations of time-averaged mean speeds. Each mean speed *vs.* temperature (16–36°C) was fit to the Arrhenius law: mean speed, $v = A\, e^{-E_a/RT}$, where $R = 8.31$ J/mol K was the gas constant, $A$ was the pre-exponential factor, $T$ was temperature, and $E_a$ was the activation energy (solid curves in B-D).

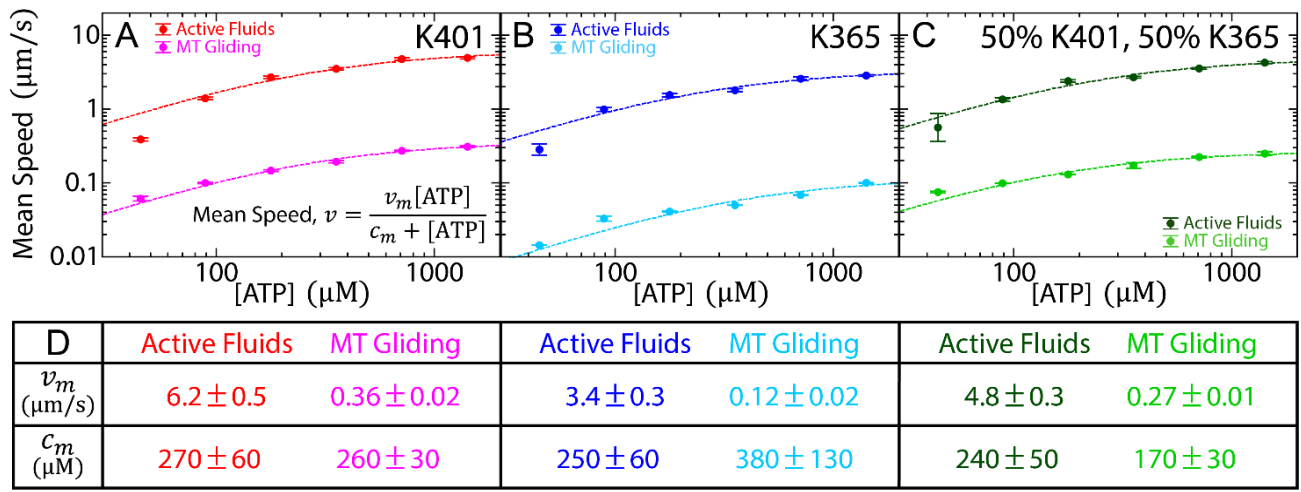

Fig. 3: Increasing ATP Concentrations Accelerated Active Fluid Flows and Microtubule Gliding. (A-C) Mean speed of active fluids and microtubule gliding at 20°C as a function of ATP concentration. Error bars represent the standard deviations of time-averaged mean speeds. Each mean speed ($v$) *vs.* ATP concentration was fitted to Michaelis–Menten kinetics: $v = v_m[\text{ATP}]/(c_m + [\text{ATP}])$, where $v_m$ was the saturation speed and $c_m$ was the ATP concentration at 50% saturation speed, $v_m/2$ (dashed curves), with $v_m$ and $c_m$ as fitting parameters. (D) Table of fitting parameters. The fitting errors were ~15% on average, demonstrating the extent of the theoretical validity in both active fluids and microtubule gliding.

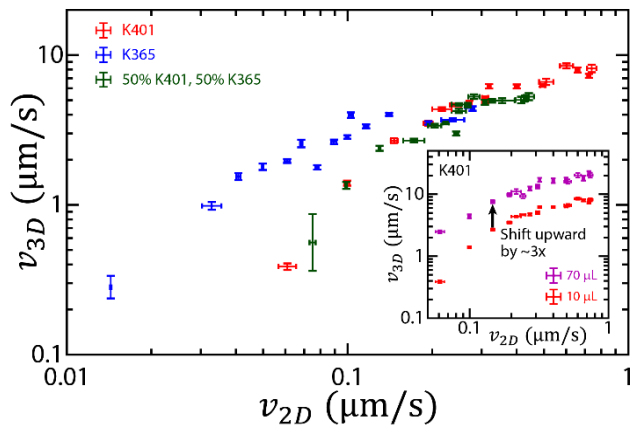

**Fig. 4**: Cross-Comparison of Mean Speeds of Microtubule-Based Active Fluid Flows in 3D ($v_{3D}$) and Kinesin-Driven Microtubule Gliding in 2D ($v_{2D}$). Microtubule gliding represented the activity of a single microtubule, whereas flows of active fluids represented the collective dynamics of multiple microtubules. Both motions were driven by either K401 (processive, red), K365 (non-processive, blue) or an equal mixture of both (dark green). Tuning up kinesin dynamics by increasing temperature or ATP concentration accelerated fluid flows and microtubule gliding. When gliding was ≳0.2 μm/s, the $v_{3D}$-$v_{2D}$ correlation was nearly identical among the explored systems, whereas below ~0.2 μm/s, the systems with different motor processivities had different $v_{3D}$-$v_{2D}$ correlations. Inset: K401-driven $v_{3D}$-$v_{2D}$ for 10- (red) and 70- (purple) μL samples. Enlarging sample volume by a factor of 7 speeded up $v_{3D}$ by ~3-fold, suggesting that volume played a role in scaling the fluid's collective dynamics. Error bars represent the standard deviations of time-averaged mean speeds.

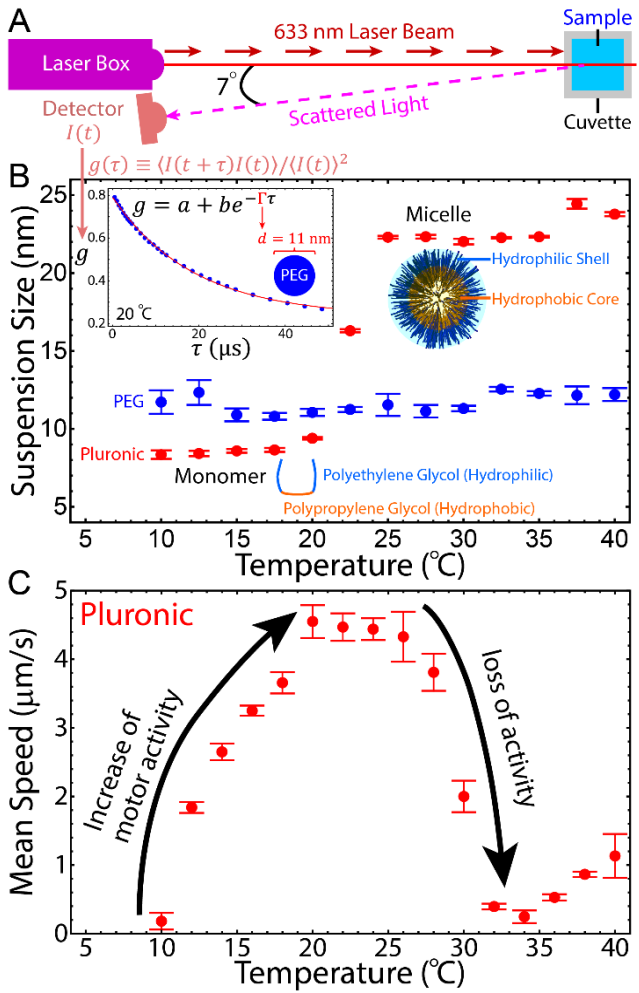

**Fig. 5**: Micelle Formation of the Pluronic Suppresses Flows of Active Fluids. (A) Schematics of measurement of suspension sizes with dynamic light scattering. The suspension sizes were measured by first exposing a sample to a laser beam. The laser beam was scattered by the suspension; the scattered light was then collected to measure autocorrelation functions $g(\tau)$, which were analyzed to determine the suspension size. (B) Suspension size as a function of temperature. Error bars represent the standard deviations of 3 measurements. PEG sizes remained nearly invariant (~11 nm) from 10 to 40°C (blue dots), whereas pluronic sizes remained at ~8 nm below 20°C, followed by an increase to ~23 nm between 20 and 25°C (red dots). The increased size suggested the formation of micelles. Inset: Intensity autocorrelation function of light scattered by 0.8% PEG in high salt M2B at 20°C. The autocorrelation function decayed exponentially; its decay rate $\Gamma$ implied PEG sizes. (C) Flow mean speeds of active fluids doped with 2% pluronic as a function of temperature. Increasing the temperature from 10 to 20°C accelerated the flows due to more active motors along with more stabilized microtubules, whereas above 20°C, pluronic started to form micelles while the fluids lost speed. Error bars represent the standard deviations of time-averaged mean speeds.

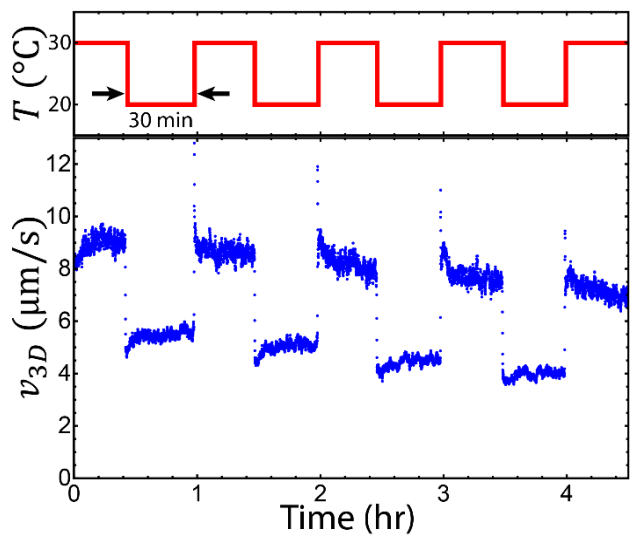

Fig. 6: Cycling Temperature Alternated Flow Speeds of Active Fluids. The flows were driven by kinesin dynamics that were tuned with temperatures in a reversing cycle. Tuning up and down kinesin dynamics periodically led flows to accelerate and decelerate accordingly, demonstrating *in-situ* control of active fluid dynamics with temperature.

**Supplementary Material for:**

**'Collective dynamics of microtubule-based 3D active fluids from single microtubules'**

Teagan E. Bate, Edward J. Jarvis, Megan E. Varney, and Kun-Ta Wu

Supplementary Information:

**Coarsening of Microtubule Bundles.** Dynamics of active fluids slowed down gradually over time (**Fig. 1D**). This decay was hypothesized to be related to structural change of active gels. To verify this hypothesis, microtubules were imaged in active gels at 20°C for 14 hr. In the beginning, microtubules were distributed mostly homogeneously, but over time, microtubules formed bundles that grew thicker (**Fig. S1**). Thickness growth enhanced image contrast, which allowed characterization of bundle coarsening by measuring the standard deviation of pixel values. The standard deviation increased with time, reflecting the growth of bundle thickness (blue in **Fig. S1C**). To examine whether the bundle coarsening caused the activity decay, the velocity field of microtubule bundle movements and their mean speeds as a function of time were measured (**Fig. S1B&C**)[11, 85]. The mean speeds decreased over time accompanied by bundle coarsening, which suggested that the coarsening of microtubule bundles slowed down their movements. The slower bundle movements led to slower active fluid flows. Fortunately, this activity decay was slow and did not impact measurement results if the observation window was short. For example, at 20°C, bundle movements slowed from ~3 to ~2 μm/s in 8 hr (**Fig. S1C**) but the decay effect was negligible during $t = 1–2$ hr, the chosen observation window.

**Characterization of Microtubule Depolymerization.** To examine the temperature dependence of the stability of GMPCPP-stabilized microtubules, microtubule lengths were monitored in microtubule gliding samples for 50 min at 10–20°C. However, gliding microtubules were motile and did not stay within a microscope field during the measurement window; therefore the sample was deprived of ATP to immobilize microtubules. The stationary microtubules were seen to shrink in lengths at 10°C whereas at 20°C the length was better preserved (**Fig. S2A**), similar to results of previous studies using taxol-treated microtubules[86]. To measure microtubule lengths, a filament tracking algorithm was used to trace microtubules in each image (pink curves), revealing microtubule length $l$ decreased with time $t$. To characterize the length decay, the time-averaged rate of the length shrinking $k$ was measured by fitting $l$ vs. $t$ to a line function, followed by extracting the line slope (inset in **Fig. S2B**). The time-averaged rates were measured for 10–13 microtubules to determine mean shrinking rate $\bar{k}$. The microtubules shrank below 16°C, implying microtubule depolymerization (**Fig. S2B**). Therefore, to ensure that microtubule depolymerization did not affect the investigation of active fluid dynamics, this analysis involved only temperature data at ≥16°C.

**Temperature-Induced Malfunction for Kinesin Clusters.** In this work, sample temperatures were varied to explore collective dynamics of active fluids. Maintaining fluid activity required kinesin to bridge pairs of microtubules while stepping toward microtubule plus ends (**Fig. 1A**). The dynamics required functional kinesin and formation of kinesin clusters. To ensure that both requirements were met, performance of kinesin clusters was examined in the explored temperature range, by adapting Böhm *et al.*'s method[35]. K401 motor clusters were pre-incubated at 20–40°C for 1 hr. The pre-incubated motors were used to prepare active fluid samples, in which time-averaged mean speeds of developed flows were measured at 20°C, and the mean speed was plotted as a function of the pre-incubation temperature (red dots in **Fig. S3A**). Mean speeds remained almost invariant (~5 μm/s) for the pre-incubation temperatures between 20 and 36°C. Above ~36°C, mean speeds dropped to ~0 μm/s, accompanied by microtubules developing into short, stationary bundles (top image in **Fig. S3B**). This result was in contrast to the cases without pre-incubation, in which microtubules developed into the long, extending bundles (**Fig. 1C**). These two observations suggested that K401 clusters were incapable of bridging and sliding pairs of microtubules after incubation at >36°C.

In this work, non-processive motor K365 was also used. To examine temperature impact on K365 motor clusters, the same experiments were repeated. The K365 clusters remained stable at 20–40°C (blue dots in **Fig. S3A**). The motor clusters were able to drive microtubules into active gels after pre-incubation at 38°C (bottom image in **Fig. S3B**), suggesting that K365 clusters were more heat resistant than K401 clusters. While the underlying mechanisms causing such a difference remained an open question, in this work with both types of motors, the data was fitted, analyzed and compared at ≲36°C, to avoid the intervention of malfunctioning motor clusters in the analysis results.

Supplementary Figures:

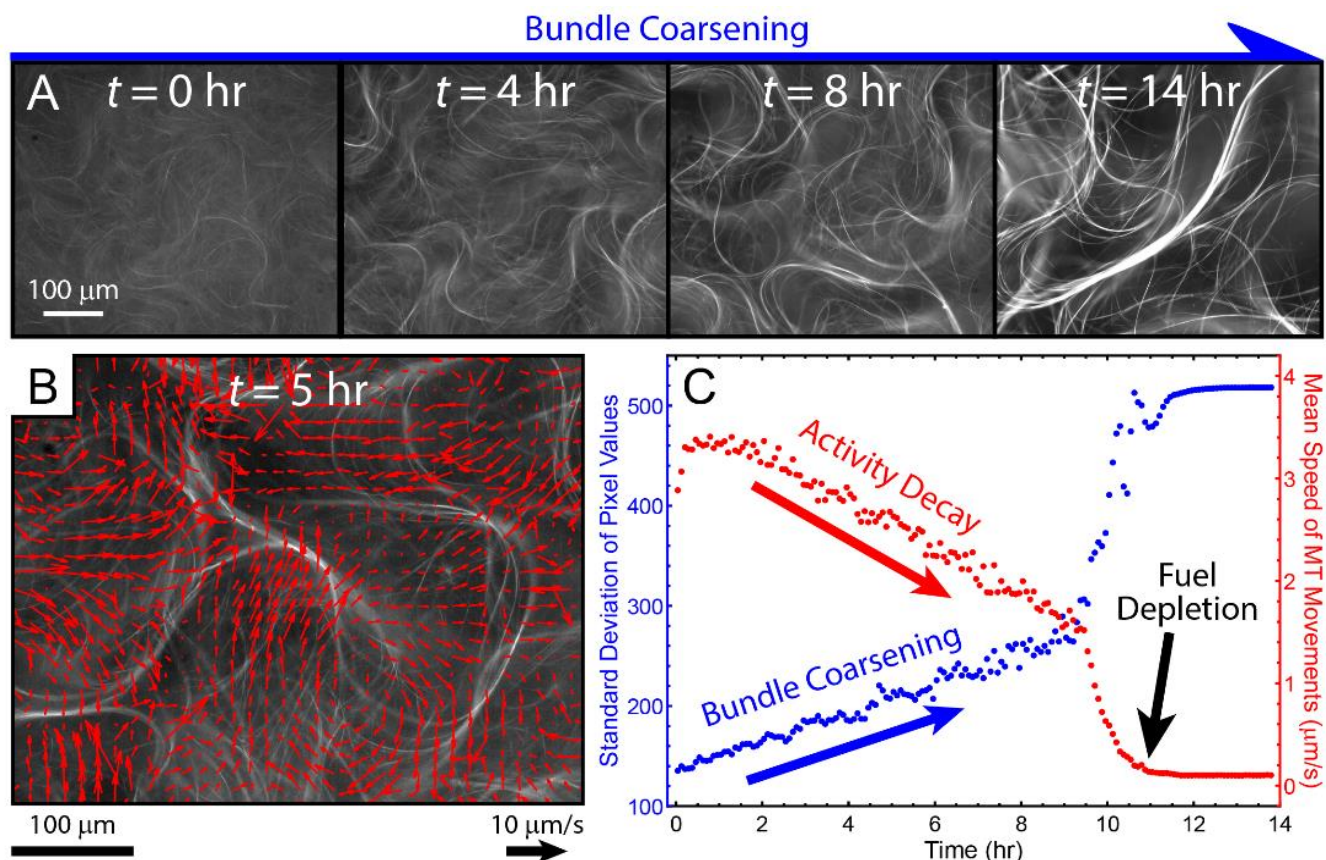

**Fig. S1:** Coarsening of Microtubule Bundles Causes Decays in Microtubule Activity. (A) Sequential images of microtubules at 20°C at $t = 0$–14 hr. The microtubules formed bundles which coarsened over time. (B) Velocity fields of movements of microtubule bundles. The velocities were measured by tracking the motion of microtubule bundles with a particle-image-velocimetry algorithm (PIVlab version 2.02)[11, 85]. (C) Standard deviation of pixel values in microtubule images compared with mean speed of microtubule movements. Bundle coarsening enhanced image contrast which increased standard deviation of the image's pixel values. The increase in the standard deviation was accompanied by the deceleration of microtubule movements, suggesting that bundle coarsening slowed down activity. Activity lasted for ~11 hr before fuel depletion.

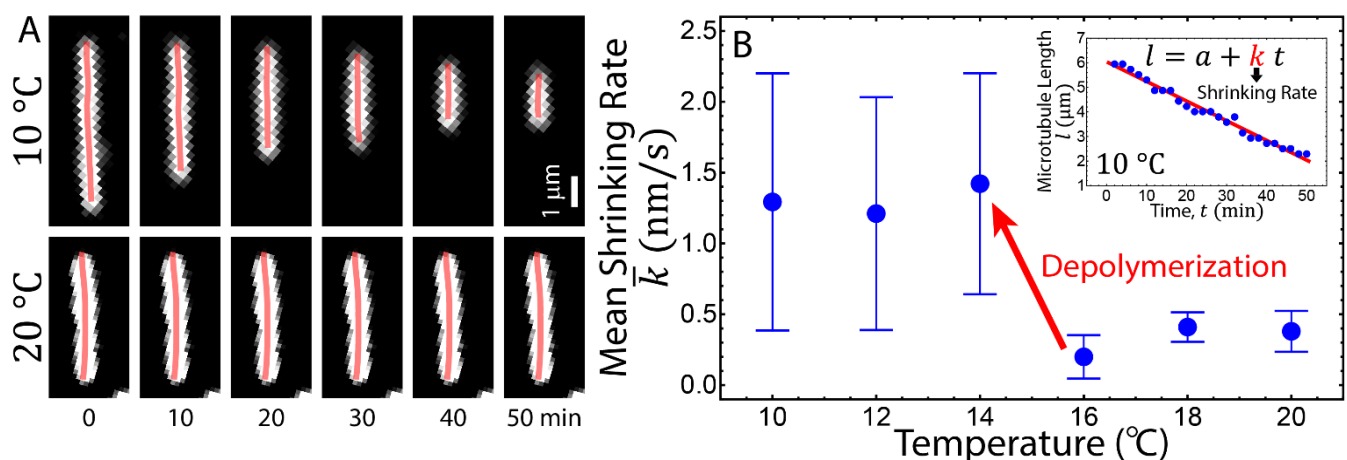

**Fig. S2:** GMPCPP-Stabilized Microtubules Depolymerize below 16°C. (A) Contrast enhanced images of GMPCPP-stabilized microtubules at 10 and 20°C (top and bottom rows) from $t = 0$ to 50 min (left to right). Each microtubule image was traced by a snake algorithm to measure the microtubule lengths (pink lines)[87]. (B) Mean shrinking rate $\bar{k}$ of microtubule length as a function of temperature. Error bars represent the standard deviations of measurements on 10–13 microtubules. Cooling microtubules below 16°C speeded up shrinking, indicating microtubule depolymerization below 16°C (red arrow). Inset: Time-averaged shrinking rate $k$ was measured by fitting the microtubule length $l$ vs. time $t$ (blue dots) to a line function $l = a + kt$ with $a$ and $k$ as fitting parameters (red line).

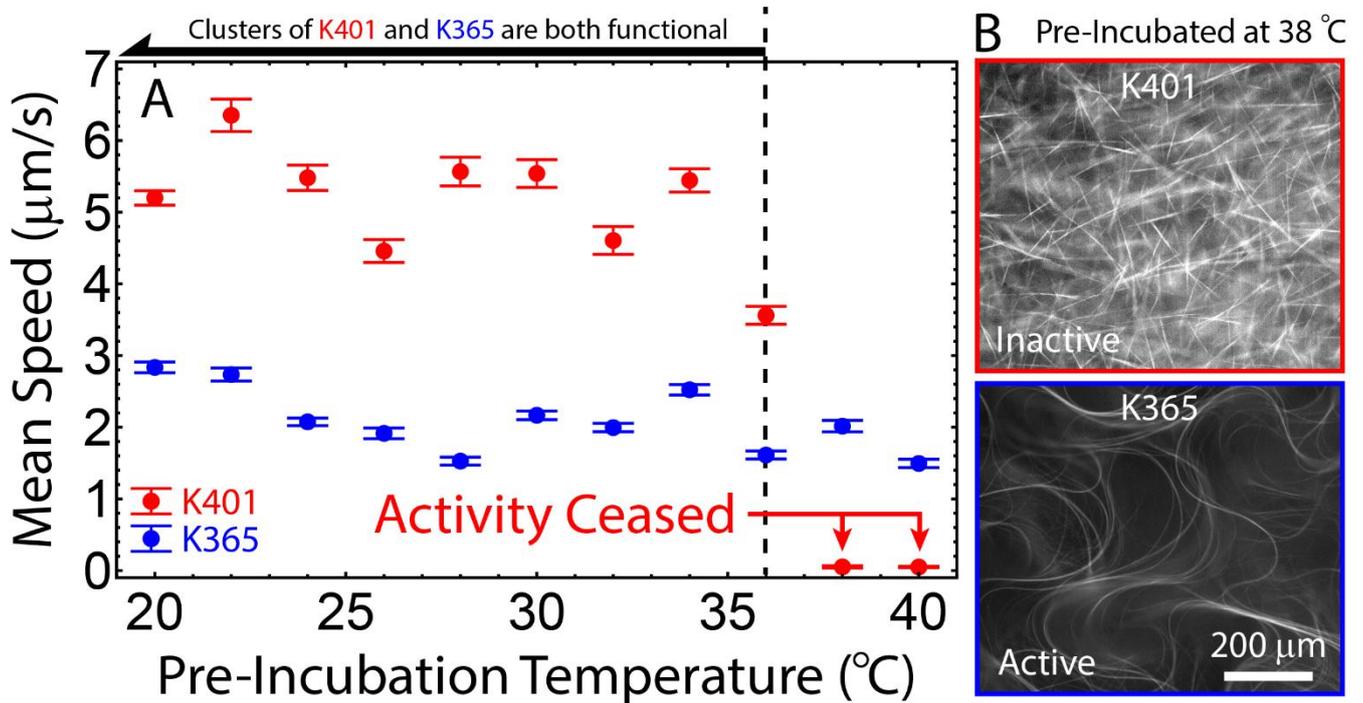

**Fig. S3:** Temperature-Induced Malfunction for Kinesin Clusters. (A) Mean speed of active fluid flows at 20°C driven by pre-incubated motors as a function of pre-incubation temperature. Error bars represent the standard deviations of time-averaged mean speeds. Pre-Incubating K401 clusters at $T \gtrsim 38°C$ ceased flows of active fluids (~0 μm/s), in contrast to the case for K365 clusters. Both types of motor clusters were functional below ~36°C. (B) Images of microtubule-based gels with clusters of K401 (top) and K365 (bottom) pre-incubated at 38°C. The pre-incubated K401 clusters failed to develop long, extensile bundles; the microtubules were stationary. In contrast, the pre-incubated K365 clusters remained capable of sustaining active gel activity.

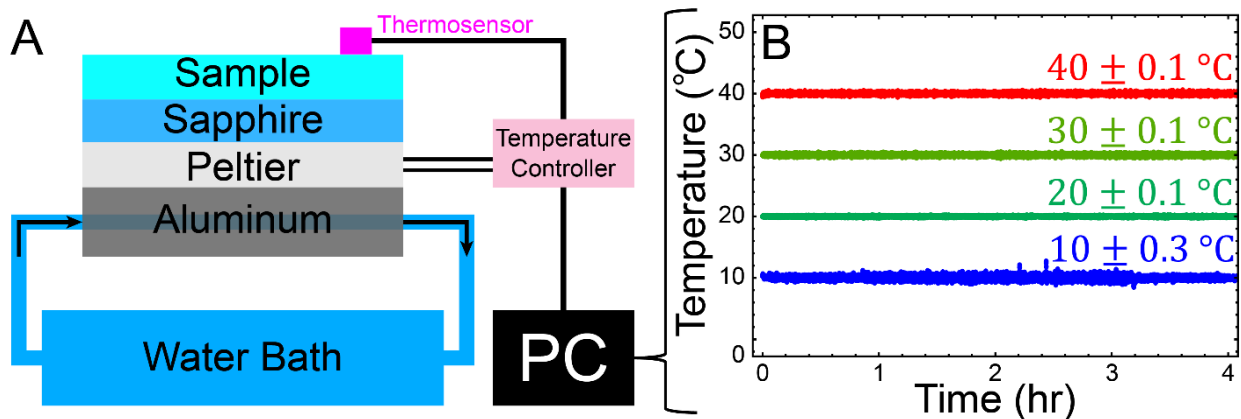

**Fig. S4**: PID-Based Temperature Control. (A) Schematic of temperature control setup. The setup sat on an aluminum stage whose temperature was regulated by internal circulating flows with a room temperature water bath. The stage supported a peltier, which cooled or heated the sample. To even the sample temperature, a sapphire was inserted between the sample and peltier. The peltier was controlled by a temperature controller that set the power of heating or cooling based on reading the sample temperature with a thermosensor. The read temperatures were recorded on an attached computer to track the temperature stability of each experiment. (B) Recorded sample temperatures for experiments at 10–40°C. Sample temperatures were controlled and monitored throughout experiments. Over the course of 4 hours, temperatures fluctuated within ±0.1–0.3°C.

Supplementary Movies:

**Movie S1**: Motion of tracers revealed flows of K401-driven active fluids at 20°C. Time stamp is hour: minute: second.

**Movie S2**: Microtubules gliding on K401-coated surfaces at 20°C. Time stamp is hour: minute: second.

**Movie S3**: Increasing temperatures from 20 to 30°C accelerated flows of active fluids, whereas decreasing from 30 to 20°C decelerated flows. Flows of active fluids could be tuned locally with temperature. Time stamp is hour: minute: second.